# What characteristics define disinformation and fake news?: review of taxonomies and definitions [English]

*Ergon Cugler de Moraes Silva*

Getulio Vargas Foundation (FGV)
University of São Paulo (USP)
São Paulo, São Paulo, Brazil

contato@ergoncugler.com
www.ergoncugler.com

*José Carlos Vaz*

University of São Paulo (USP)
Getulio Vargas Foundation (FGV)
São Paulo, São Paulo, Brazil

vaz@usp.br
www.getip.net.br

**Abstract**

What characteristics define disinformation and fake news? To address this research question, this Technical Note provides a comprehensive analysis of disinformation and fake news, synthesizing 46 definitions and highlighting four key points addressing their fundamental characteristics. Adopting the "Prisma 2020" method, five search sets with the Boolean operator 'AND' were selected in both Portuguese and English, which were applied across four databases, resulting in 237 reviewed articles. Following a meticulous analysis, relevant articles were identified and included, while duplicates and inaccessible documents were excluded. It points to 'disinformation as information that is totally or partially false, crafted by a sender with the aim of misleading, with opportunistic content designed to manipulate reality, being amplified by individual characteristics of the receiver in their interpretation and by contextual factors in which they are embedded'. This Technical Note seeks to contribute to an understanding of the phenomenon of disinformation that includes the contextual dimension, obtaining as fundamental elements of analysis: I.) Sender; II.) Content; III.) Receiver; and IV.) Environment.

## 1. Introduction

This research is the result of a broader study that listed 46 definitions of disinformation and fake news, all detailed in Table 01, further ahead. In summary, we can highlight four points that seek to answer the following question: "What characteristics define disinformation and fake news?": **I.) Sender:** There is a set of authors for whom the characteristic factors of disinformation or fake news stem from the intentionality of its sender (or agent, be it producer or propagator), there may be content that is not necessarily false in its entirety but has been manipulated to dispute some opinion; **II.) Content:** There is a set of authors for whom, regardless of intentionality, what characterizes disinformation or fake news is the aspect in which such false information manifests itself, which can be organized into a series of typologies, ranging from the most humorous to the most conspiratorial, these factors being potentializers of their efficiency; **III.) Receiver:** There is a set of authors who interpret disinformation or fake news to the extent that someone - the 'interpreter', 'receiver' - is victimized by it, and this incidence is aggravated by characteristics or interpretive factors of the receiver, these being cognitive, psychological, ideological, or mobilizing knowledge, skills, thinking styles, emotions, values, and may lead to confusion, doubt, or assertion of trust due to group belonging or identity; and **IV.) Environment:** There is a set of authors who base how the environment can not only shape disinformation or fake news but also characterize it, both by a series of factors such as demographic, social, political, economic, among others; and by the so-called "echo chambers" that serve as incubators in a feedback dynamic to characterize information as false information. That is, the context in which information circulates as capable of shaping its harmful or falsified dimension, resulting from the interaction among the participants of such an environment, reconfiguring such information.

Therefore, amidst disinformation as a 'multifaceted and polysemic phenomenon', a synthesis encompassing the four emphases is pointed out: **'<u>disinformation as information that is totally or partially false, crafted by a sender with the aim of misleading, with opportunistic content designed to manipulate reality, being amplified by individual characteristics of the receiver in their interpretation and by contextual factors in which they are embedded</u>'**.

## 2. Materials and methods

This research adopted, for its realization, the "Prisma 2020" method (Page et al., 2021), which is a set of evidence-based items for systematic review reports. To do so, 05 sets with the Boolean operator 'AND' were selected, in Portuguese: "fake news" AND "review"; "desinformação" AND "review" and in English: "fake news" AND "review"; "misinformation" AND "review"; "disinformation" AND "review".

These sets were searched in 04 databases, Oasis BR Ibict; SciELO International (Scientific Electronic Library Online); Scopus; and Web of Science, returning 237 results with the keywords in the title or abstract. After the survey, eligible articles for analysis were identified, applying inclusion criteria and excluding those that only address the citation of misinformation in some topic, do not propose theories, classifications, or taxonomies, are duplicates, or have inaccessible documents. These steps are presented in Figure 01:

**Figure 01.** "Prisma 2020 Framework" model for systematic literature review:

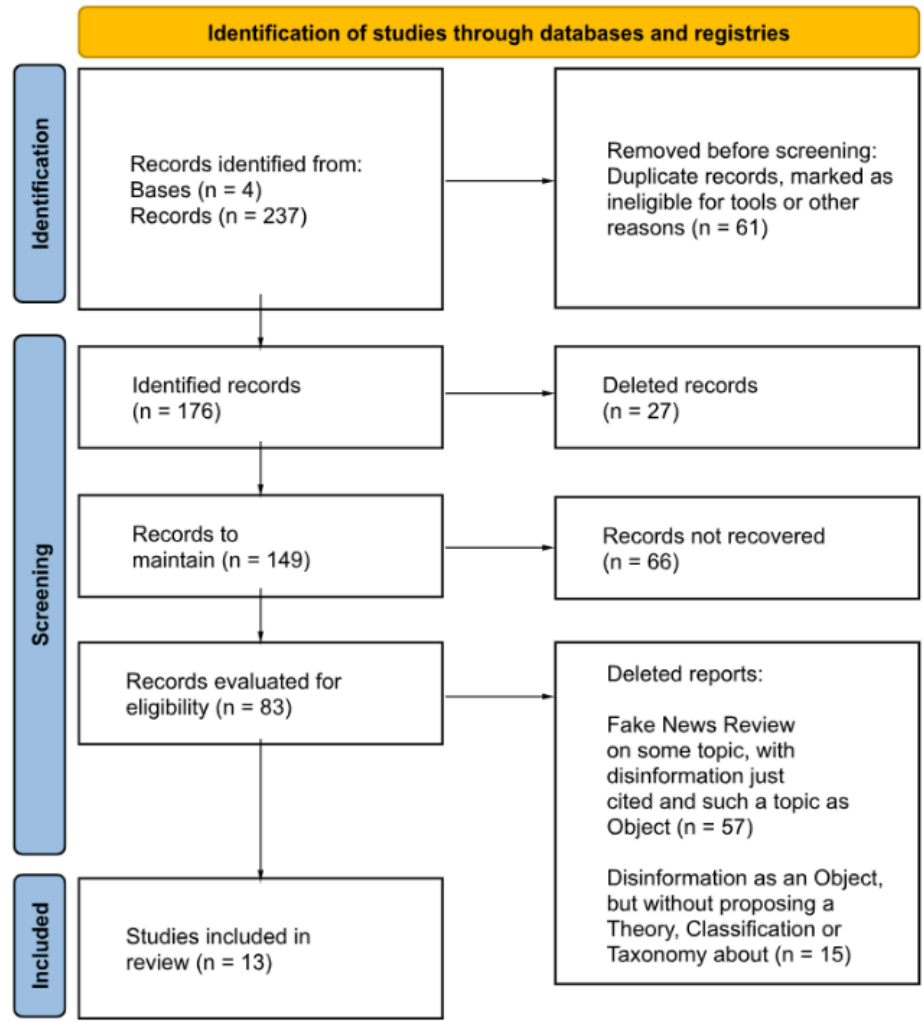

Source: Authors (2024).

## 3. Exploring the results

**Table 01.** Concepts and classifications in the literature on disinformation:

| Author(s) | Review source(s) | Emphasis | Main contributions |
|---|---|---|---|
| Wardle e Derakhshan (2017) | Abad (2018) | - | Three elements of "disinformation" are listed, being a.) "agents"; b.) "message"; and c.) "interpreter". |
| Baptista & Gradim (2020)* | Baptista & Gradim (2020) | Content | Systematizes that a.) "emotional content"; b.) "heuristic persuasion"; c.) "imitation of the journalistic format"; d.) "clickbait"; and e.) "images" in addition to dimensions of characterization of "fake news", are also potentializers. |
| Damstra et al. (2021)* | Damstra et al. (2021) | Content | They point out that "fake news", or "intentional deception" content, is characterized by one or more of the following characteristics, in addition to false content, being: a.) use and presence of emotions; b.) difficulty of verifiability; c.) long and attractive headlines; d.) lexical diversity / ("type-token ratio"); e.) excessive capitalization; f.) excessive use of pronouns; g.) shorter article length but longer tweets; h.) words and informal language, in addition to hate speech; and i.) less use of punctuation in the text, but more in the headline. |
| Jang. et al. (2018) | Di Domenico et al. (2021) | Content | Tweets about real news showed greater breadth and less depth than tweets about "fake news". |
| Jaster e Lanius (2018) | Baptista & Gradim (2020) | Content | They argue that "fake news" does not necessarily depend on intentionality, but on being false. |
| Rastogi & Bansal (2022)* | Rastogi & Bansal (2022) | Content | They review typologies of so-called "misinfomation", with aspects of the content of such misleading information characterizing its denomination, which may be: a.) fake news; b.) conspiracy theories; c.) "hoax"; d.) rumors; e.) "clickbait"; and f.) satirical news. |
| Resende et al. (2019); Souza et al. (2018); Vosoughi et al. (2018) | Forster et al. (2021) | Content | "Fake news" as content that mimics some elements of professional journalistic texts, such as layout and macrostructuring, as well as appealing elements that make a text more attractive to the digital universe. |
| Rochlin (2017) | Di Domenico et al. (2021) | Content | "Fake news" as "a knowingly false headline and story written and published on a website designed to look like a real news website and spread across social media." |
| Tandoc, Lim & Ling (2018) | Di Domenico et al. (2021); Estrada-Cuzcano et al. (2020) | Content | They identified six ways in which previous studies operationalize "fake news", being: a.) satire; b.) parody; c.) manufacturing; d.) manipulation; e.) advertising; and f.) advertising. |
| Argo et al. (2005); Jagatic et al. (2007); Jun et al. (2007); Sterret et al. (2018) | Forster et al. (2021) | Environment | In terms of social influence and trust, it suggests that the mere perception of social presence can reduce the vigilance of individuals in relation to statements, characterizing the effectiveness of "fake news" due to the environment in which it is inserted. |
| Bessi et al. (2016) | Di Domenico et al. (2021) | Environment | It points out how the environment in which information is available can create "echo chambers". Furthermore, the authors show that user comment patterns are accurate predictors for the formation of "echo chambers". |

| Author(s) | Review source(s) | Emphasis | Main contributions |
|---|---|---|---|
| Bryanov & Vziatysheva (2021)* | Bryanov & Vziatysheva (2021) | Environment | They point out how "confirmation bias" can be reinforced in a narrative that is shaped according to how it receives interactions in a "herd effect". That is, the environment in which such content exists and its dynamics of interaction with users as a characteristic of disinformation in the making. |
| Chowdhury et al. (2021)* | Chowdhury et al. (2021) | Environment | They point out seven factors to characterize "fake news", factors: a.) demographic; b.) intrapersonal; c.) interpersonal/social; d.) professional/experimental; e.) related to the source of information; f.) related to risk communication; and g.) related to government/authority, being contextual elements. |
| Di Domenico et al. (2021)* | Di Domenico et al. (2021) | Environment | "Fake news" will affect consumers and businesses differently depending on the context. "Fake news" will change consumers' knowledge, attitudes and behaviors about brands/products in everyday life. |
| Nan et al. (2022)* | Nan et al. (2022) | Environment | They list 11 demographic and contextual factors that predict susceptibility to "disinformation" in health, including: a.) age; b.) sex; c.) education; d.) income; e.) race (specifically, racial minority, African American, Hispanic American); f.) geographic region (specifically, living in an urban area); and g.) employment. |
| Nan et al. (2022)* | Nan et al. (2022) | Environment | They list 11 behavioral factors that predict susceptibility to "disinformation" in health, including: a.) use of social media; b.) use of mainstream media; c.) conservative use of the media; d.) use of health professionals or scientists as sources of information; e.) use of friends or family as sources of information; f.) general use of online/digital media; and g.) difficulty in seeing doctors |
| Rastogi & Bansal (2022)* | Rastogi & Bansal (2022) | Environment | "Biased", or biased information, in a kind of biased "echo chamber", such as "4chan's/pol/board", an imageboard website in English, in which users publish anonymously and constitute a lively informational dynamic, in which this can be distorted by the context and, therefore, biased. |
| Visentin et al. (2019) | Di Domenico et al. (2021) | Environment | They highlight that "fake news" can produce different consequences that spread in different contexts — covering not only trust and attitudes, but also behavioral consequences, such as purchase intention, word-of-mouth recommendation and intention to access consumption. |
| Allcott & Gentzkow (2017) | Di Domenico et al. (2021) | Sender | Intentionally "wrong" or "fake" news is produced to make money and/or promote ideologies. |
| Brito e Pinheiro (2015) | Abad (2018) | Sender | "Disinformation" as an instrument with the aim of disinforming someone and leading to deception, with intent. |
| Demo (2000) | Abad (2018) | Sender | "Disinformation" as "provision of informational products of low cultural level", for "imbecilization". |
| Di Domenico et al. (2021)* | Di Domenico et al. (2021) | Sender | "Fake news" is created to pursue financial and/or ideological objectives and is materialized through manufactured legitimacy to allow the content to go viral. |
| Lazer et al. (2018) | Di Domenico et al. (2021) | Sender | "Fake news" is news that has been "fabricated", but presented as if from legitimate sources, and promoted on social media to deceive the public for ideological and/or financial gain. |
| Lewis & Marwick (2017); Marwick (2018); Vorderer et al. (2004); Coleman (2014) | Baptista & Gradim (2020) | Sender | "Disinformation" as an instrument for "the ambition to attract attention, denigrate the image of a political candidate, impose a certain ideological belief or encourage some type of behavior from users". Or, "simply for fun or to create chaos." Being characterized based on the objective of its issuer. |

| Author(s) | Review source(s) | Emphasis | Main contributions |
|---|---|---|---|
| Martens et al. (2018) | Di Domenico et al. (2021) | Sender | "Fake news" as "disinformation" which includes all forms of false, inaccurate, or misleading information designed, presented, and promoted to intentionally or for profit cause public harm (e.g., commercial click-bait). |
| Meneses (2018) | Baptista & Gradim (2020) | Sender | "False news" differs from "fake news", the first of which does not involve intentionality or fraud. |
| Özgöbek & Gulla (2017) | Di Domenico et al. (2021) | Sender | "Fake news" is intentionally manufactured to be misleading and may prove to be false. |
| Rastogi & Bansal (2022)* | Rastogi & Bansal (2022) | Sender | "Propaganda" as a special instance of fabricated stories that aim to harm a certain party, such as the narrative dispute surrounding BlackLivesMatter, or the airstrikes in Syria in 2018. In this aspect, the intentionality of the manufacturer characterizes the informational dispute. |
| Rini (2017) | Di Domenico et al. (2021) | Sender | "Fake news" as news that purports to describe events in the real world, usually by imitating the reporting conventions of traditional media, but is known by its creators to be significantly false and is broadcast with the aim of being widely rebroadcast and of misleading at least part of your audience |
| Wardle e Derakhshan (2017) | Abad (2018); Forster et al. (2021); Rastogi & Bansal (2022) | Sender | Three variations are classified, being "Disinformation", false and harmful information; "Misinformation", false and not necessarily harmful information; and "Malinformation", harmful information that is not necessarily false. |
| Zhang & Ghorbani (2020) | Di Domenico et al. (2021) | Sender | "Fake news" refers to the types of false stories or news that are published and distributed primarily on the Internet in order to purposely mislead, mislead, or lure readers for financial, political, or other gain. |
| Britt et al. (2019) | Di Domenico et al. (2021) | Receiver | They analyze belief and memory biases that play a role in the formation of belief in "fake news". |
| Bronstein et al. (2019) | Bryanov & Vziatysheva (2021); Di Domenico et al. (2021) | Receiver | They investigate how "delusional ideation", "dogmatism" and "religious fundamentalism" directly correlated with belief in "fake news", but did not correlate with belief in real news. |
| Budd & Stewart (2018); Petrucco & Ferranti (2017) | Alonso Varela & Saraiva Cruz (2020) | Receiver | Two steps are segmented to identify "disinformation", being "search for information" and "evaluation of information". It is pointed out that "disinformation", therefore, is not the result of a mere absence of information, but depends on the critical evaluation of the receiver (or its absence). |
| Chua & Banerjee (2018); Dubois & Blank (2018); Estabel et al. (2020); Pennycook et al. (2015) | Forster et al. (2021) | Receiver | It points to evidence that prior exposure to "false information", its familiarity, makes it more likely to be judged as true, even for subjects who are not aware of the repetition. Or, for false information to have an impact, in addition to attracting the receiver, it needs to be processed as true by at least a portion of those who receive it. In addition to the level of literacy as a characteristic of effectiveness. |
| Damstra et al. (2021)* | Damstra et al. (2021) | Receiver | They point to a characteristic of (Hyper-)partisan bias, generally with a right-wing ideological orientation. |
| Di Domenico et al. (2021)* | Di Domenico et al. (2021) | Receiver | Regardless of its nature and intention, a promoter seeks to obtain a set of responses from consumers. These cognitive, emotional and behavioral responses determine individuals' belief in fake news, which, in turn, enables the spread of "fake news". |

| Author(s) | Review source(s) | Emphasis | Main contributions |
|---|---|---|---|
| Estrada-Cuzcano et al. (2020)* | Estrada-Cuzcano et al. (2020) | Receiver | It points out how excessive access to large amounts of data and information has led to saturation and, consequently, displaced, in many cases, the veracity of information and the critical attitude towards the information that is received or obtained. It reinforces the reflection that "disinformation" is not just a lack of information from the receiver. |
| Gelfert (2018) | Di Domenico et al. (2021) | Receiver | "Fake news" as cases of deliberate presentation of typically false or misleading claims as news, where these are misleading by design, (...) systemic characteristics of the sources and channels through which fake news propagates and therefore manipulates (...) pre-existing cognitive biases and heuristics of consumers. |
| Harmonjones & Mills (2017); Horne & Adali (2017); Jang & Kim (2018); Knobloch-Westerwick et al. (2020) | Forster et al. (2021) | Receiver | "Confirmation bias", "cognitive dissonance" and "heuristic processing", which characterize "fake news". |
| Jones-Jang et al. (2019) | Bryanov & Vziatysheva (2021) | Receiver | It identifies how adherence to "fake news" is "significantly associated with information literacy", with disinformation characterized by the interpretative limitation of the content receiver. |
| Nan et al. (2022)* | Nan et al. (2022) | Receiver | They list 46 psychological factors that predict susceptibility to "disinformation" in health, including: a.) knowledge/skills; b.) thinking style; c.) trust; d.) emotion; e.) value; and f.) group identity. |
| Nehmy & Paim (1998); Aquino (2007) | Abad (2018) | Receiver | "Disinformation" as "the individual's state of ignorance regarding the knowledge that would be relevant to them". |
| Rapp & Salovich (2018) | Di Domenico et al. (2021) | Receiver | Exposure to "inaccurate information" creates at least three problems with understanding: a.) confusion; b.) doubt; and c.) trust, all of which shape aspects of "fake news" in its relationship with the consumer. |
| Sadiq & Saji (2022)* | Sadiq & Saji (2022) | Receiver | Points out how: a.) anxiety; b.) source ambiguity; c.) reliability; d.) content ambiguity; d.) personal involvement; e.) social ties; f.) confirmation bias; g.) attractiveness; h.) illiteracy; i.) ease of sharing options; and j.) attachment to the device are determining variables of "disinformation". |
| Tello (1998) | Peña (2022) | Receiver | It characterizes "news" as "information, but not just any information, but that which was not known before it was made explicit and whose narration may be of interest to a wide audience, even without any connection with the event". |

Source: Own elaboration. *From authors reviewed in their publication.

## 4. References

Abad, A. C. C. (2018). Uma revisão de literatura sobre desinformação. *Trabalho de Conclusão de Curso*. http://pantheon.ufrj.br/handle/11422/11856

Alonso Varela, L., & Saraiva Cruz, I. (2020). Búsqueda y evaluación de información: Dos competencias necesarias en el contexto de las fake news. *Palabra clave*, *9*(2), 90–90. https://doi.org/10.24215/18539912e090

Baptista, J. P., & Gradim, A. (2020). Understanding Fake News Consumption: A Review. *Social Sciences*, *9*(10), Art. 10. https://doi.org/10.3390/socsci9100185

Bryanov, K., & Vziatysheva, V. (2021). Determinants of individuals' belief in fake news: A scoping review determinants of belief in fake news. *PLOS ONE*, *16*(6), e0253717. https://doi.org/10.1371/journal.pone.0253717

Chowdhury, N., Khalid, A., & Turin, T. C. (2021). Understanding misinformation infodemic during public health emergencies due to large-scale disease outbreaks: A rapid review. *Journal of Public Health*, 1–21. https://doi.org/10.1007/s10389-021-01565-3

Damstra, A., Boomgaarden, H. G., Broda, E., Lindgren, E., Strömbäck, J., Tsfati, Y., & Vliegenthart, R. (2021). What Does Fake Look Like? A Review of the Literature on Intentional Deception in the News and on Social Media. *Journalism Studies*, *22*(14), 1947–1963. https://doi.org/10.1080/1461670X.2021.1979423

Di Domenico, G., Sit, J., Ishizaka, A., & Nunan, D. (2021). Fake news, social media and marketing: A systematic review. *Journal of Business Research*, *124*, 329–341. https://doi.org/10.1016/j.jbusres.2020.11.037

Estrada-Cuzcano, A., Alfaro-Mendives, K., Saavedra-Vásquez, V., Estrada-Cuzcano, A., Alfaro-Mendives, K., & Saavedra-Vásquez, V. (2020). Disinformation y Misinformation, Posverdad y Fake News: Precisiones conceptuales, diferencias, similitudes y yuxtaposiciones. *Información, cultura y sociedad*, *42*, 93–106. https://doi.org/10.34096/ics.i42.7427

Forster, R., Carvalho, R. M. de, Filgueiras, A., & Avila, E. (2021). *Fake News: O Que É, Como Se Faz E Por Que Funciona?* SciELO Preprints. https://doi.org/10.1590/SciELOPreprints.3294

Genesini, S. (2018). A pós-verdade é uma notícia falsa. *Revista USP*, *116*, Art. 116. https://doi.org/10.11606/issn.2316-9036.v0i116p45-58

Lazer, D. M. J., Baum, M. A., Benkler, Y., Berinsky, A. J., Greenhill, K. M., Menczer, F., Metzger, M. J., Nyhan, B., Pennycook, G., Rothschild, D., Schudson, M., Sloman, S. A., Sunstein, C. R., Thorson, E. A., Watts, D. J., & Zittrain, J. L. (2018). The science of fake news. *Science*, *359*(6380), 1094–1096. https://doi.org/10.1126/science.aao2998

Nan, X., Wang, Y., & Thier, K. (2022). Why do people believe health misinformation and who is at risk? A systematic review of individual differences in susceptibility to health misinformation. *Social Science & Medicine*, *314*, 115398. https://doi.org/10.1016/j.socscimed.2022.115398

Page, M. J *et al*. (2021). Prisma 2020 explanation and elaboration: Updated guidance and exemplars for reporting systematic reviews. BMJ, 372, n160. https://doi.org/10.1136/bmj.n160

Peña, R. de la. (2022). Noticias falsas en tiempos de la posverdad. *Revista mexicana de opinión pública*, *33*, 89–102. https://doi.org/10.22201/fcpys.24484911e.2022.33.82237

Rastogi, S., & Bansal, D. (2022). A review on fake news detection 3T's: Typology, time of detection,

taxonomies. *International Journal of Information Security*, 1–36. https://doi.org/10.1007/s10207-022-00625-3

Sadiq, M. T., & Saji, M. K. (2022). The disaster of misinformation: A review of research in social media. *International Journal of Data Science and Analytics*, *13*(4), 271–285. https://doi.org/10.1007/s41060-022-00311-6

Tandoc, E. C., Lim, Z. W., & Ling, R. (2018). Defining "Fake News". *Digital Journalism*, *6*(2), 137–153. https://doi.org/10.1080/21670811.2017.1360143

Wardle, C. (2016). *6 types of misinformation circulated this election season*. Columbia Journalism Review. https://www.cjr.org/tow_center/6_types_election_fake_news.php

## 5. Authors biography

**Ergon Cugler de Moraes Silva** has a Master's degree in Public Administration and Government (FGV), Postgraduate MBA in Data Science & Analytics (USP) and Bachelor's degree in Public Policy Management (USP). He is associated with the Bureaucracy Studies Center (NEB FGV), collaborates with the Interdisciplinary Observatory of Public Policies (OIPP USP), with the Study Group on Technology and Innovations in Public Management (GETIP USP) with the Monitor of Political Debate in the Digital Environment (Monitor USP) and with the Working Group on Strategy, Data and Sovereignty of the Study and Research Group on International Security of the Institute of International Relations of the University of Brasília (GEPSI UnB). He is also a researcher at the Brazilian Institute of Information in Science and Technology (IBICT), where he works for the Federal Government on strategies against disinformation. São Paulo, São Paulo, Brazil. Web site: https://ergoncugler.com/.

**José Carlos Vaz** is a Professor at the University of São Paulo - School of Arts, Sciences, and Humanities, in the undergraduate and graduate courses in Public Policy Management. Vice-president of the Administrative Council of the Institute Pólis. Coordinator of GETIP - Study Group on Technology and Innovation in Public Management. Bachelor's degree in Administration from the University of São Paulo (1986), Master's degree in Public Administration from Fundação Getulio Vargas EAESP (1995), and a Ph.D. in Business Administration - Information Systems from Fundação Getúlio Vargas EAESP (2003). Has experience in the field of Public Administration, focusing mainly on the following themes: social and political aspects of information technology use (digital participation, open government data, social control of governments, e-government, technology public policies), public management (state and government capabilities, innovations in public management, logistics, strategic planning), and urban and municipal issues (urban processes and dynamics, urban mobility, municipal management, local development).

# Quais características definem a desinformação e as fake news?: levantamento de taxonomias e definições [Português]

*Ergon Cugler de Moraes Silva*

Fundação Getulio Vargas (FGV)
Universidade de São Paulo (USP)
São Paulo, São Paulo, Brasil

contato@ergoncugler.com
www.ergoncugler.com

*José Carlos Vaz*

Universidade de São Paulo (USP)
Fundação Getulio Vargas (FGV)
São Paulo, São Paulo, Brasil

vaz@usp.br
www.getip.net.br

**Resumo**

Quais características definem a desinformação e as fake news? Para responder a essa pergunta de pesquisa, esta Nota Técnica oferece uma análise abrangente sobre desinformação e fake news, sintetizando 46 definições e destacando quatro pontos-chave que abordam suas características fundamentais. Adotando o método "Prisma 2020", foram selecionados cinco conjuntos de busca com o operador booleano 'AND' em português e em inglês, os quais foram aplicados em quatro bases de dados, resultando em 237 artigos revisados. Após uma análise criteriosa, foram identificados e incluídos os artigos pertinentes, excluindo duplicatas e documentos inacessíveis. Aponta-se para a 'desinformação como informação total ou parcialmente falsa, elaborada por um emissor com objetivo de induzir ao erro, com conteúdo oportunista e desenhado para manipular a realidade, sendo potencializada por características individuais do receptor em sua interpretação e por fatores contextuais em que se está inserido'. Esta Nota Técnica busca contribuir para uma compreensão sobre o fenômeno da desinformação que inclua a dimensão contextual, obtendo como elementos fundamentais de análise: I.) Emissor; II.) Conteúdo; III.) Receptor; e IV.) Ambiente.

## 1. Introdução

Este levantamento é o resultado de um estudo mais abrangente que listou 46 definições de desinformação e notícias falsas, todas detalhadas no Quadro 01, mais adiante. Em síntese, podemos destacar quatro pontos que buscam responder à seguinte pergunta: "Quais características definem a desinformação e as fake news?": **I.) Emissor:** Há um conjunto de autores para os quais os fatores característicos de uma desinformação ou fake news partem da intencionalidade de seu emissor (ou agente, seja produtor ou propagador), podendo existir um conteúdo que não seja necessariamente falso em sua totalidade, mas que veio a ser manipulado para disputar alguma opinião; **II.) Conteúdo:** Há um conjunto de autores para os quais, independente da intencionalidade, o que caracteriza uma desinformação ou fake news está no aspecto em que tal informação falsa se manifesta, podendo organizar-se em uma série de tipologias, que vão desde as mais humorísticas, até as mais conspiratórias, sendo tais fatores potencializadores de sua eficiência; **III.) Receptor:** Há um conjunto de autores que interpretam uma desinformação ou fake news na medida que alguém - o 'intérprete', 'receptor' - é vitimado por tal, sendo essa incidência agravada por características ou fatores interpretativos do receptor intérprete, sendo estes cognitivos, psicológicos, ideológicos ou que mobilizem conhecimentos, habilidades, estilos de pensamento, emoções, valores e podem levar à confusão, dúvida ou afirmação de confiança por pertencimento ou identidade de grupo; e **IV.) Ambiente:** Há um conjunto de autores que embasam como o ambiente pode não apenas modelar uma desinformação ou fake news, mas também caracterizá-la, tanto por uma série de fatores como demográficos, sociais, políticos, econômicos, dentre outros; quanto pelas chamadas "câmaras de eco" que servem como incubadoras em dinâmica retroalimentada para caracterizar uma informação como então uma falsa informação. Isto é, o contexto em que uma informação é circulada como capaz de modelar sua dimensão danosa ou falseada, resultante da interação entre os partícipes de tal ambiente, capazes de reconfigurar tal informação.

Aponta-se, portanto, em meio à desinformação como 'fenômeno multifacetado e polissêmico', em uma síntese que abarque as quatro ênfases, sendo: a **'desinformação como informação total ou parcialmente falsa, elaborada por um emissor com objetivo de induzir ao erro, com conteúdo oportunista e desenhado para manipular a realidade, sendo potencializada por características individuais do receptor em sua interpretação e por fatores contextuais em que se está inserido'**.

## 2. Materiais e métodos

Esta pesquisa adotou, para sua realização, o método "Prisma 2020" (Page et al., 2021), sendo um conjunto de itens baseado em evidências para relatórios em revisões sistemáticas. Para tal, foram selecionados 05 conjuntos com operador booleano 'AND', sendo em Português: "fake news" AND "revisão"; "desinformação" AND "revisão" e em Inglês: "fake news" AND "review"; "misinformation" AND "review"; "disinformation" AND "review".

Tais conjuntos foram buscados em 04 bases, sendo Oasis BR Ibict; SciELO International (Scientific Electronic Library Online); Scopus; e Web of Science, retornando 237 resultados com as palavras-chaves no título ou resumo. Após o levantamento, foram identificados artigos elegíveis para análise, aplicando critérios de inclusão e excluindo aqueles que tratam apenas da citação de desinformação em algum tema, não propõem teorias, classificações ou taxonomias, são duplicatas ou possuem documentos inacessíveis. Tais etapas são apresentadas na Figura 01:

**Figura 01.** Modelo "Prisma 2020 Fluxogram" para revisão sistemática de literatura:

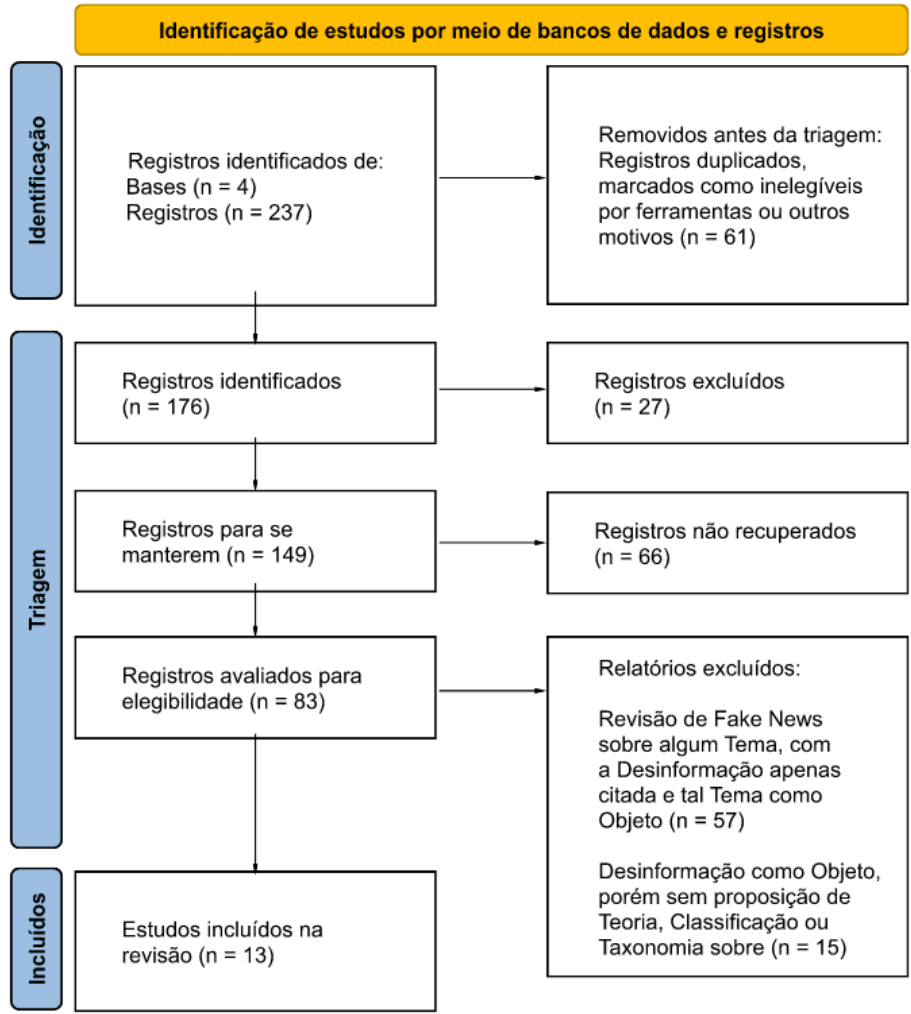

Fonte: Elaboração própria (2024).

## 3. Explorando os resultados

**Quadro 01.** Conceitos e classificações na literatura sobre desinformação:

| Autor(es) | Fonte(s) da revisão | Ênfase | Principais contribuições |
|---|---|---|---|
| Wardle e Derakhshan (2017) | Abad (2018) | - | São elencados três elementos de uma "desinformação", sendo a.) "agentes"; b.) "mensagem"; e c.) "intérprete". |
| Baptista & Gradim (2020)* | Baptista & Gradim (2020) | Conteúdo | Sistematizam que a.) "conteúdo emocional"; b.) "persuasão heurística"; c.) "imitação do formato jornalístico"; d.) "clickbait"; e e.) "imagens" além de dimensões de caracterização das "fake news", também são potencializadoras. |
| Damstra et al. (2021)* | Damstra et al. (2021) | Conteúdo | Apontam que uma "fake news", ou um conteúdo de "engano intencional" caracteriza-se por uma ou mais seguintes características, além do conteúdo falso, sendo: a.) uso e presença de emoções; b.) dificuldade de verificabilidade; c.) manchetes longas e atraentes; d.) diversidade lexical / ("type-token ratio"); e.) capitalização excessiva; f.) uso excessivo de pronomes; g.) comprimento de artigos menores, mas tweets maiores; h.) palavras e linguagem informal, além de discurso de ódio; e i.) menor uso de pontuação no texto, mas maior na manchete. |
| Jang. et al. (2018) | Di Domenico et al. (2021) | Conteúdo | Tweets sobre notícias reais mostraram maior amplitude e menor profundidade do que tweets sobre "fake news". |
| Jaster e Lanius (2018) | Baptista & Gradim (2020) | Conteúdo | Defendem que "fake news" não dependem, necessariamente, de intencionalidade, mas de ser falsa. |
| Rastogi & Bansal (2022)* | Rastogi & Bansal (2022) | Conteúdo | Revisam tipologias das chamadas "misinfomation", sendo que aspectos do conteúdo de tal informação enganosa caracterizam sua denominação, podendo ser: a.) fake news; b.) teorias de conspiração; c.) "hoax"; d.) rumores; e.) "clickbait"; e f.) notícias satíricas. |
| Resende et al. (2019); Souza et al. (2018); Vosoughi et al. (2018) | Forster et al. (2021) | Conteúdo | "Fake news" como conteúdos que mimetizam alguns elementos de textos jornalísticos profissionais, como o layout e a macroestruturação, além de elementos apelativos que configuram um texto mais atraente ao universo digital. |
| Rochlin (2017) | Di Domenico et al. (2021) | Conteúdo | "Fake news" como "uma manchete e uma história sabidamente falsas, escritas e publicadas em um site projetado para se parecer com um site de notícias real e espalhadas pelas mídias sociais". |
| Tandoc, Lim & Ling (2018) | Di Domenico et al. (2021); Estrada-Cuzcano et al. (2020) | Conteúdo | Identificaram seis maneiras pelas quais estudos anteriores operacionalizam "fake news", sendo: a.) sátira; b.) paródia; c.) fabricação; d.) manipulação; e.) propaganda; e f.) publicidade. |
| Argo et al. (2005); Jagatic et al. (2007); Jun et al. (2007); Sterret et al. (2018) | Forster et al. (2021) | Ambiente | Em influência social e confiança, sugere que a mera percepção de presença social pode reduzir a vigilância de indivíduos em relação a afirmações, caracterizando a eficácia de uma "fake news" pelo Ambiente de sua inserção. |
| Bessi et al. (2016) | Di Domenico et al. (2021) | Ambiente | Aponta como o ambiente em que a informação está disposta pode criar "câmaras de eco". Além disso, os autores mostram que os padrões de comentários dos usuários são preditores precisos para a formação de "câmaras de eco". |

| Autor(es) | Fonte(s) da revisão | Ênfase | Principais contribuições |
|---|---|---|---|
| Bryanov & Vziatysheva (2021)* | Bryanov & Vziatysheva (2021) | Ambiente | Apontam como o "viés de confirmação" pode ser reforçado em uma narrativa que se molda conforme recebe interações em "efeito manada". Isto é, o ambiente em que tal conteúdo está e sua dinâmica de interação com usuários como caracterizadora de uma desinformação em construção. |
| Chowdhury et al. (2021)* | Chowdhury et al. (2021) | Ambiente | Apontam sete fatores para caracterizar as "fake news", fatores: a.) demográficos; b.) intrapessoais; c.) interpessoais/sociais; d.) profissionais/experimentais; e.) relacionados à fonte de informação; f.) relacionados à comunicação de risco; e g.) relacionados ao governo/autoridade, sendo elementos contextuais. |
| Di Domenico et al. (2021)* | Di Domenico et al. (2021) | Ambiente | As "fake news" afetarão os consumidores e as empresas de maneira diferente, dependendo do contexto. "Fake news" mudarão o conhecimento, atitudes e comportamentos dos consumidores sobre marcas/produtos no cotidiano. |
| Nan et al. (2022)* | Nan et al. (2022) | Ambiente | Listam 11 fatores demográficos como contextuais que predizem a suscetibilidade à "desinformação" em saúde, passando por: a.) idade; b.) sexo; c.) educação; d.) renda; e.) raça (especificamente, minoria racial, afro-americano, hispano-americano); f.) região geográfica (especificamente, vivendo em uma área urbana); e g.) emprego. |
| Nan et al. (2022)* | Nan et al. (2022) | Ambiente | Listam 11 fatores comportamentais que predizem a suscetibilidade à "desinformação" em saúde, passando por: a.) uso de mídias sociais; b.) uso da grande mídia; c.) uso conservador da mídia; d.) uso de profissionais de saúde ou cientistas como fontes de informação; e.) uso de amigos ou familiares como fontes de informação; f.) uso geral de mídia online/digital; e g.) dificuldade em consultar médicos |
| Rastogi & Bansal (2022)* | Rastogi & Bansal (2022) | Ambiente | "Biased", ou informação enviesada, em uma espécie de "câmara de eco" tendenciosa, como o "4chan's/pol/board", um website imageboard em inglês, no qual os usuários publicam anonimamente e constituem uma dinâmica viva informacional, na qual esta pode ser distorcida em meio ao contexto e, portanto, enviesada. |
| Visentin et al. (2019) | Di Domenico et al. (2021) | Ambiente | Destacam que "fake news" podem produzir diferentes consequências que se espalham em diferentes contextos — abrangendo não apenas confiança e atitudes, mas também consequências comportamentais, como intenção de compra, indicação boca a boca e intenção de acesso ao consumo. |
| Allcott & Gentzkow (2017) | Di Domenico et al. (2021) | Emissor | Notícias intencionalmente "erradas" ou "fakes" são produzidas para ganhar dinheiro e/ou promover ideologias. |
| Brito e Pinheiro (2015) | Abad (2018) | Emissor | "Desinformação" como instrumento com objetivo de desinformar alguém e conduzir ao engano, com dolo. |
| Demo (2000) | Abad (2018) | Emissor | "Desinformação" como "fornecimento de produtos informacionais de baixo nível cultural", para "imbecilização". |
| Di Domenico et al. (2021)* | Di Domenico et al. (2021) | Emissor | As "fake news" são criadas para perseguir objetivos financeiros e/ou ideológicos e são materializadas por meio de uma legitimidade fabricada para permitir a viralidade do conteúdo. |
| Lazer et al. (2018) | Di Domenico et al. (2021) | Emissor | "Fake news" são notícias que foram "fabricadas", mas apresentadas como se fossem de fontes legítimas, e promovidas nas mídias sociais para enganar o público para obter ganhos ideológicos e/ou financeiros. |
| Lewis & Marwick (2017); Marwick (2018); Vorderer et al. (2004); Coleman (2014) | Baptista & Gradim (2020) | Emissor | "Desinformação" como instrumento para "a ambição de chamar a atenção, denegrir a imagem de um candidato político, impor uma determinada crença ideológica ou encorajar algum tipo de comportamento dos usuários". Ou, "simplesmente por diversão ou para criar o caos". Sendo caracterizada a partir do objetivo de seu emissor. |

| Autor(es) | Fonte(s) da revisão | Ênfase | Principais contribuições |
|---|---|---|---|
| Martens et al. (2018) | Di Domenico et al. (2021) | Emissor | "Fake news" como "desinformação" que inclui todas as formas de informações falsas, imprecisas ou enganosas projetadas, apresentadas e promovidas para causar danos públicos intencionalmente ou com fins lucrativos (por exemplo, click-bait comercial). |
| Meneses (2018) | Baptista & Gradim (2020) | Emissor | Difere "false news" de "fake news", sendo que a primeira não conta com intencionalidade, ou dolo. |
| Özgöbek & Gulla (2017) | Di Domenico et al. (2021) | Emissor | "Fake news" são intencionalmente fabricadas para serem enganosas e podem provar que são falsas. |
| Rastogi & Bansal (2022)* | Rastogi & Bansal (2022) | Emissor | "Propaganda" como uma instância especial de histórias fabricadas que visam prejudicar uma determinada parte, como disputa narrativa em torno do BlackLivesMatter, ou os ataques aéreos na Síria em 2018. Neste aspecto, a intencionalidade do fabricante caracteriza a disputa informacional. |
| Rini (2017) | Di Domenico et al. (2021) | Emissor | "Fake news" como aquela que pretende descrever eventos no mundo real, geralmente imitando as convenções da reportagem da mídia tradicional, mas é conhecida por seus criadores como significativamente falsa e é transmitida com o objetivo de ser amplamente retransmitida e de enganar pelo menos parte de seu público |
| Wardle e Derakhshan (2017) | Abad (2018); Forster et al. (2021); Rastogi & Bansal (2022) | Emissor | São classificadas três variações, sendo "Disinformation", informação falsa e nociva; "Misinformation", informação falsa e não necessariamente nociva; e "Malinformation", informação nociva e não necessariamente falsa. |
| Zhang & Ghorbani (2020) | Di Domenico et al. (2021) | Emissor | "Fake news" refere-se aos tipos de histórias ou notícias falsas que são publicadas e distribuídas principalmente na Internet, a fim de enganar, enganar ou atrair leitores propositadamente para ganhos financeiros, políticos ou outros. |
| Britt et al. (2019) | Di Domenico et al. (2021) | Receptor | Analisam vieses de crença e memória que desempenham papel na formação da crença em "fake news". |
| Bronstein et al. (2019) | Bryanov & Vziatysheva (2021); Di Domenico et al. (2021) | Receptor | Investigam como a "ideação delirante", o "dogmatismo" e o "fundamentalismo religioso" se correlacionaram diretamente com a crença em "fake news", mas que não se correlacionaram com a crença em notícias reais. |
| Budd & Stewart (2018); Petrucco & Ferranti (2017) | Alonso Varela & Saraiva Cruz (2020) | Receptor | São segmentadas duas etapas para identificar uma "desinformação", sendo "busca de informação" e "avaliação da informação". Aponta-se que a "desinformação", portanto, não é resultado de mera ausência informacional, mas depende de avaliação crítica do receptor (ou de sua ausência). |
| Chua & Banerjee (2018); Dubois & Blank (2018); Estabel et al. (2020); Pennycook et al. (2015) | Forster et al. (2021) | Receptor | Aponta evidência de que a exposição prévia a uma "informação falsa", sua familiaridade, torna mais provável ela seja julgada como verdadeira, mesmo para sujeitos que não estão conscientes da repetição. Ou, ainda, A informação falsa, para que tenha impacto, além de atrair o receptor, precisa ser processada como verdadeira por pelo menos uma parte daqueles que a recebem. Além do grau de letramento como caracterizador da eficácia. |
| Damstra et al. (2021)* | Damstra et al. (2021) | Receptor | Apontam para uma característica de (Hiper-)viés partidário, geralmente com uma orientação ideológica de direita. |
| Di Domenico et al. (2021)* | Di Domenico et al. (2021) | Receptor | Independentemente de sua natureza e intenção, um divulgador busca obter um conjunto de respostas dos consumidores. Essas respostas cognitivas, emocionais e comportamentais determinam a crença dos indivíduos em notícias falsas, o que, por sua vez, possibilita a disseminação de "fake news". |

| Autor(es) | Fonte(s) da revisão | Ênfase | Principais contribuições |
|---|---|---|---|
| Estrada-Cuzcano et al. (2020)* | Estrada-Cuzcano et al. (2020) | Receptor | Aponta como o acesso excessivo a grandes quantidades de dados e informação tem levado à saturação e, consequentemente, deslocado, em muitos casos, a veracidade da informação e a atitude crítica face à informação que se recebe ou obtém. Reforça a reflexão de que "desinformação" não é só ausência de informação do receptor. |
| Gelfert (2018) | Di Domenico et al. (2021) | Receptor | "Fake news" como casos de apresentação deliberada de alegações tipicamente falsas ou enganosas como notícias, onde estas são enganosas por design, (...) características sistêmicas das fontes e canais pelos quais notícias falsas se propagam e, portanto, manipulam (...) vieses cognitivos e heurísticas pré-existentes dos consumidores. |
| Harmonjones & Mills (2017); Horne & Adali (2017); Jang & Kim (2018); Knobloch-Westerwick et al. (2020) | Forster et al. (2021) | Receptor | "Viés de confirmação", "dissonância cognitiva" e "processamento heurístico", caracterizadores de "fake news". |
| Jones-Jang et al. (2019) | Bryanov & Vziatysheva (2021) | Receptor | Identifica como a adesão às "fake news" é "significativamente associada à alfabetização informacional", sendo que a desinformação caracteriza-se pela limitação interpretativa do receptor do conteúdo. |
| Nan et al. (2022)* | Nan et al. (2022) | Receptor | Listam 46 fatores psicológicos que predizem a suscetibilidade à "desinformação" em saúde, passando por: a.) conhecimento/habilidades; b.) estilo de pensamento; c.) confiança; d.) emoção; e.) valor; e f.) identidade do grupo. |
| Nehmy & Paim (1998); Aquino (2007) | Abad (2018) | Receptor | "Desinformação" como "estado de ignorância do indivíduo em relação ao conhecimento que lhe seria relevante". |
| Rapp & Salovich (2018) | Di Domenico et al. (2021) | Receptor | A exposição a "informações imprecisas" cria pelo menos três problemas de compreensão: a.) confusão; b.) dúvida; e c.) confiança, todos os quais moldam os aspectos das "fake news" em sua relação com o consumidor. |
| Sadiq & Saji (2022)* | Sadiq & Saji (2022) | Receptor | Aponta como: a.) ansiedade; b.) ambiguidade de fonte; c.) confiabilidade; d.) ambiguidade de conteúdo; d.) envolvimento pessoal; e.) laços sociais; f.) viés de confirmação; g.) atratividade; h.) analfabetismo; i.) facilidade de opções de compartilhamento; e j.) apego ao dispositivo são variáveis determinantes da "desinformação". |
| Tello (1998) | Peña (2022) | Receptor | Caracteriza "news" como "uma informação, mas não qualquer informação, mas aquela que não se sabia antes de ser explicitada e cuja narração pode interessar a um público amplo, inclusive sem ligação com o acontecimento". |

Fonte: Elaboração própria. *A partir de autores revisados em sua publicação.

## 4. Referências

Abad, A. C. C. (2018). Uma revisão de literatura sobre desinformação. *Trabalho de Conclusão de Curso*. http://pantheon.ufrj.br/handle/11422/11856

Alonso Varela, L., & Saraiva Cruz, I. (2020). Búsqueda y evaluación de información: Dos competencias necesarias en el contexto de las fake news. *Palabra clave*, *9*(2), 90–90. https://doi.org/10.24215/18539912e090

Baptista, J. P., & Gradim, A. (2020). Understanding Fake News Consumption: A Review. *Social Sciences*, *9*(10), Art. 10. https://doi.org/10.3390/socsci9100185

Bryanov, K., & Vziatysheva, V. (2021). Determinants of individuals' belief in fake news: A scoping review determinants of belief in fake news. *PLOS ONE*, *16*(6), e0253717. https://doi.org/10.1371/journal.pone.0253717

Chowdhury, N., Khalid, A., & Turin, T. C. (2021). Understanding misinformation infodemic during public health emergencies due to large-scale disease outbreaks: A rapid review. *Journal of Public Health*, 1–21. https://doi.org/10.1007/s10389-021-01565-3

Damstra, A., Boomgaarden, H. G., Broda, E., Lindgren, E., Strömbäck, J., Tsfati, Y., & Vliegenthart, R. (2021). What Does Fake Look Like? A Review of the Literature on Intentional Deception in the News and on Social Media. *Journalism Studies*, *22*(14), 1947–1963. https://doi.org/10.1080/1461670X.2021.1979423

Di Domenico, G., Sit, J., Ishizaka, A., & Nunan, D. (2021). Fake news, social media and marketing: A systematic review. *Journal of Business Research*, *124*, 329–341. https://doi.org/10.1016/j.jbusres.2020.11.037

Estrada-Cuzcano, A., Alfaro-Mendives, K., Saavedra-Vásquez, V., Estrada-Cuzcano, A., Alfaro-Mendives, K., & Saavedra-Vásquez, V. (2020). Disinformation y Misinformation, Posverdad y Fake News: Precisiones conceptuales, diferencias, similitudes y yuxtaposiciones. *Información, cultura y sociedad*, *42*, 93–106. https://doi.org/10.34096/ics.i42.7427

Forster, R., Carvalho, R. M. de, Filgueiras, A., & Avila, E. (2021). *Fake News: O Que É, Como Se Faz E Por Que Funciona?* SciELO Preprints. https://doi.org/10.1590/SciELOPreprints.3294

Genesini, S. (2018). A pós-verdade é uma notícia falsa. *Revista USP*, *116*, Art. 116. https://doi.org/10.11606/issn.2316-9036.v0i116p45-58

Lazer, D. M. J., Baum, M. A., Benkler, Y., Berinsky, A. J., Greenhill, K. M., Menczer, F., Metzger, M. J., Nyhan, B., Pennycook, G., Rothschild, D., Schudson, M., Sloman, S. A., Sunstein, C. R., Thorson, E. A., Watts, D. J., & Zittrain, J. L. (2018). The science of fake news. *Science*, *359*(6380), 1094–1096. https://doi.org/10.1126/science.aao2998

Nan, X., Wang, Y., & Thier, K. (2022). Why do people believe health misinformation and who is at risk? A systematic review of individual differences in susceptibility to health misinformation. *Social Science & Medicine*, *314*, 115398. https://doi.org/10.1016/j.socscimed.2022.115398

Page, M. J *et al*. (2021). Prisma 2020 explanation and elaboration: Updated guidance and exemplars for reporting systematic reviews. BMJ, 372, n160. https://doi.org/10.1136/bmj.n160

Peña, R. de la. (2022). Noticias falsas en tiempos de la posverdad. *Revista mexicana de opinión pública*, *33*, 89–102. https://doi.org/10.22201/fcpys.24484911e.2022.33.82237

Rastogi, S., & Bansal, D. (2022). A review on fake news detection 3T's: Typology, time of detection,

taxonomies. *International Journal of Information Security*, 1–36. https://doi.org/10.1007/s10207-022-00625-3

Sadiq, M. T., & Saji, M. K. (2022). The disaster of misinformation: A review of research in social media. *International Journal of Data Science and Analytics*, *13*(4), 271–285. https://doi.org/10.1007/s41060-022-00311-6

Tandoc, E. C., Lim, Z. W., & Ling, R. (2018). Defining "Fake News". *Digital Journalism*, *6*(2), 137–153. https://doi.org/10.1080/21670811.2017.1360143

Wardle, C. (2016). *6 types of misinformation circulated this election season*. Columbia Journalism Review. https://www.cjr.org/tow_center/6_types_election_fake_news.php

## 5. Biografia dos autores

**Ergon Cugler de Moraes Silva** possui mestrado em Administração Pública e Governo (FGV), MBA pós-graduação em Ciência de Dados e Análise (USP) e bacharelado em Gestão de Políticas Públicas (USP). Ele está associado ao Núcleo de Estudos da Burocracia (NEB FGV), colabora com o Observatório Interdisciplinar de Políticas Públicas (OIPP USP), com o Grupo de Estudos em Tecnologia e Inovações na Gestão Pública (GETIP USP), com o Monitor de Debate Político no Meio Digital (Monitor USP) e com o Grupo de Trabalho sobre Estratégia, Dados e Soberania do Grupo de Estudo e Pesquisa sobre Segurança Internacional do Instituto de Relações Internacionais da Universidade de Brasília (GEPSI UnB). É também pesquisador no Instituto Brasileiro de Informação em Ciência e Tecnologia (IBICT), onde trabalha para o Governo Federal em estratégias contra a desinformação. São Paulo, São Paulo, Brasil. Site: https://ergoncugler.com/.

**José Carlos Vaz** é Professor da Universidade de São Paulo - Escola de Artes, Ciências e Humanidades, nos cursos de graduação e de pós-graduação em Gestão de Políticas Públicas. Vice-presidente do Conselho Administrativo do Instituto Pólis. Coordenador do GETIP - Grupo de Estudos em Tecnologia e Inovação na Gestão Pública. Graduação em Administração pela Universidade de São Paulo (1986), Mestrado em Administração Pública pela Fundação Getúlio Vargas SP (1995) e doutorado em Administração de Empresas - Sistemas de Informação pela Fundação Getúlio Vargas SP (2003). Tem experiência na área de Administração Pública, atuando principalmente nos seguintes temas: aspectos sociais e políticos do uso da tecnologia de informação (participação digital, dados governamentais abertos, controle social dos governos, governo eletrônico, políticas públicas de tecnologia), gestão pública (capacidades estatais e de governo, inovações em gestão pública, logística, planejamento estratégico) e questões urbanas e municipais (processos e dinâmicas urbanas, mobilidade urbana, gestão municipal, desenvolvimento local).